\documentclass{aa}
\usepackage{graphics}

\usepackage{natbib}
\usepackage{graphicx}
\usepackage{amssymb}
\begin{document}

\title{Spatially resolved X-ray emission of EQ Pegasi}

\author{J. Robrade, J.-U. Ness, J.H.M.M. Schmitt}
\institute{
Hamburger Sternwarte, Universit\" at Hamburg, Gojenbergsweg 112,
D-21029 Hamburg, Germany}

\authorrunning{Robrade et al.}
\titlerunning{XMM observation of EQ Peg}
\offprints{J. Robrade}
\mail{jrobrade@hs.uni-hamburg.de}
\date{Received September 11, 2003; accepted October 1, 2003}

\abstract{
We present an analysis of an XMM-Newton observation of the M dwarf binary EQ
Pegasi with a special focus on the the spatial structure of the X-ray emission
and the analysis of light curves.
Making use of data obtained with EPIC (European Photon Imaging Camera) we were
for the first time able to spatially
resolve the two components in X-rays and to study the light curves
of the individual components of the EQ Peg system.
During the observation a series of moderate flares was detected,
where it was possible to identify the respective flaring component.
}
\maketitle

\section{Introduction}
\label{intro}
X-ray observations with the {\it Einstein Observatory} and ROSAT
have shown the ubiquitous occurrence of coronae around most classes of stars.
ROSAT studies of volume-limited complete samples of cool stars in the immediate
solar neighborhood have shown coronal formation around late-type cool dwarf
stars with outer convection zones to be universal;
all stars investigated with sufficient sensitivity were found to be
surrounded by X-ray emitting coronae \citep[][]{schmitt95, schmitt03}.
Interestingly, fully convective M dwarfs have also been found
to be very active with frequent flares.

EQ Peg is a nearby (6.25\,pc) visual binary (period $\sim 180$\,yr, separation
5.2\arcsec) consisting of two M dwarfs of spectral type M3.5 and M4.5. It was first observed
photoelectrically to flare by \cite{roques54}, and
\cite{owen72} found both components of the system to be
flare stars.

EQ Peg has been observed at radio, optical, EUV, and X-ray wavelengths.
Observations in the optical focused on the flare nature of EQ Peg and marked
emission line variability during photometric quiescence was found
\citep{bopp74} as well as frequent optical flares on both components
\citep{rod78}.
A VLA map of EQ Peg at 6\,cm was presented by \cite{topka82}. They
resolved both components and interpreted the radio emission as ``quiescent''
since they found it unlikely that both components flared
at the same time. The radio emission was confined to each component and \cite{topka82}
concluded that radio production mechanisms do not depend on binary interaction
(which is plausible due to the separation of $\sim 25$\,AU).

EQ Peg was observed by all major previous imaging X-ray missions and again found to
flare frequently. EQ Peg was observed with the {\it Einstein Observatory}
\citep{vai81} and is contained in the ESS (Einstein Slew Survey) \citep{elvis92}.
EXOSAT detected an intense long duration flare during a coordinated observation
with the VLA \citep{palla86}; a detailed modelling of these flares and the underlying
physical properties is presented by \cite{poletto88}.  EQ Peg was also detected in
the ROSAT all-sky survey \citep{huensch99} and
rapid flaring was simultaneously
observed at optical and X-ray wavelength with MEKASPEC and ROSAT \citep{katsova02}, where the source
brightened in X-rays by a factor of $\sim$ 15.
A coordinated VLA, optical, EUVE, and RXTE monitoring of EQ Peg was carried out
by \cite{gagne98}. They found a classic stellar flare with a rapid impulsive
phase (radio burst) followed by rapid chromospheric heating and cooling (U-band)
and more gradual coronal cooling (X-ray and extreme-UV). In addition they found
atypical flares with either highly polarized emission with no counterparts
at shorter wavelengths or moderately polarized flares that often have
shorter-wavelength counterparts.

EQ Peg was also observed with XMM-Newton. In Sect.~\ref{anal} we describe the observations
and the methods used for data
analysis. Here we focus on the data from the EPIC instruments
in order to obtain spatial and temporal information on the two components of EQ Peg.
In Sect.~\ref{results} we present the results followed by a summary and 
discussion in Sect.~\ref{summ}.

\section{Observation and data analysis}
\label{anal}

EQ Peg A/B (V=10.32\,mag/12.4\,mag) was observed on 2000 July 9 (MJD=51734) with XMM-Newton.
The 15\,ksec observation of EQ Peg (see Tab. \ref{obs}) provided useful data in
all EPIC (European Photon Imaging Camera) detectors. 
The EPIC instrument consists of three CCD cameras with two different types of CCD design, resp. two MOS 
(Metal Oxide Semi-conductor CCDs) and one PN (pn CCDs), providing imaging and spectroscopy.
The EPIC cameras offer the possibility to perform extremely sensitive imaging observations 
over the telescope's field of view of 30 $\arcmin$ and in the energy range from 0.15 to 15 keV 
with good angular and moderate spectral resolution. A detailed description of the XMM instruments 
can be found in \cite{xmm}.
All EPIC instruments (MOS/PN) operated in the full frame
mode with the thick filter inserted. Unless otherwise indicated we used for our purposes the
full energy bandpass of the EPIC instruments, resp. 0.15\,keV to 12.0/15.0\,keV.

\begin{table}[!ht]
\caption{\label{obs}Observation log of EQ Peg}
{\scriptsize
\begin{tabular}{lcc}\hline
 Instrument (Mode) & Duration (s) & Obs-time\\
 MOS (FF,thick F.) & 14600 & 2000-07-09T11:39:13-15:42:32 \\
 PN  (FF,thick F.) & 12410 & 2000-07-09T12:20:16-15:47:05 \\
\end{tabular}
}
\end{table}

\begin{figure}[ht]
 \resizebox{\hsize}{!}{\rotatebox{270}{\includegraphics{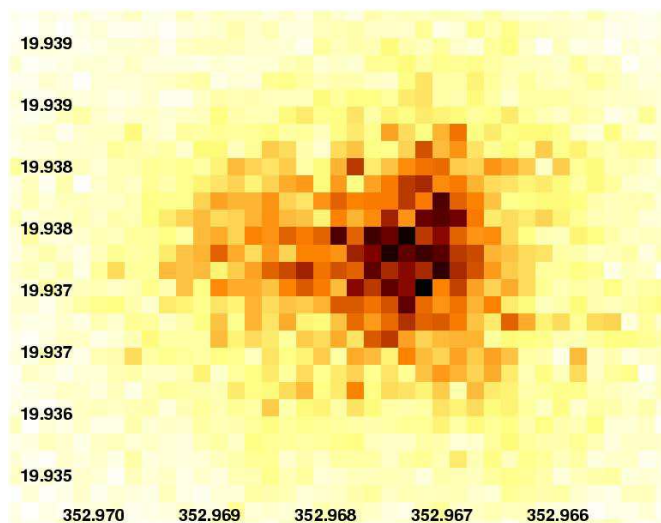}}}
\caption{\label{image}Image of the EQ Peg system (MOS1) with linear brightness
scaling. The image is elongated, suggesting the presence of two components. 
Analysis shows that the X-ray brighter component is EQ Peg~A.}
\end{figure}

The data were reduced with the standard XMM-Newton Science Analysis System (SAS)
software, version 5.4.1. Light curves and images were produced with standard
SAS tools and standard selection criteria were applied for filtering the data.
In Fig.~\ref{image} we show the image obtained with the MOS1 detector.  The
image looks elongated and it is reasonable to assume the elongation is due to emission from
both components of EQ Peg. The image elongation can
be seen clearly in MOS1, but not in MOS2 and PN due to the triangular
shape of the point spread function (PSF) for MOS2 and the bigger pixel size in PN.

For a quantitative analysis of the MOS1 image
we developed a fitting procedure applicable to the measured event
distribution in order to confirm the detection of the two components of EQ Peg and to
determine accurate source positions and count rates. For this procedure
we optimize a set of parameters describing the modelled event distribution on the sky plane.
The modelled event distribution is constructed on the basis of the PSF
\citep{psf}, which is composed of a King component plus background
\begin{equation}
\label{model}
{\rm PSF} = \frac{A}{(1+(\frac{r}{r_{c}})^2)^{\alpha}} +{\rm BKG}
\end{equation}
with the core radius $r_c$, the slope $\alpha$ (model parameters), the
distance to the peak position $r$, and the amplitude $A$ (source parameters).
The background flux BKG was determined from source free regions in the detector
and is kept as a fixed model parameter. We place two such King components near
the positions of the two sources and apply an optimization algorithm seeking
best-fit values for the amplitudes $A_1$ and $A_2$ and the positions $r_1=(x_1,y_1)$ and
$r_2=(x_2,y_2)$ of each component. We thus determine the model counts in the spatial
bin $(i,j)$
\begin{equation}
{\bf n_{i,j}}=\frac{A_1}{(1+(\frac{r_1-r_{i,j}}{r_{c}})^2)^{\alpha}}
+ \frac{A_2}{(1+(\frac{r_2-r_{i,j}}{r_{c}})^2)^{\alpha}} +{\rm BKG}
\end{equation}
and apply Powell's algorithm \citep{powell}, a robust multi-dimensional 
minimizing routine, in order to minimize the likelihood function
\begin{equation}
{\cal L} = -2\sum_{i,j}\log({\bf n}_{i,j}*{\bf c}_{i,j}-{\bf n}_{i,j})
\end{equation}
with the measured counts ${\bf c}_{i,j}$ and model counts ${\bf n}_{i,j}$.
The model counts ${\bf n}_{i,j}$ are constrained to conserve the total
number of counts, i.e., $\sum_{i,j}{\bf n}_{i,j}=\sum_{i,j}{\bf c}_{i,j}$.\\

The mean PSF-parameters ($r_c$ and $\alpha$) as determined by in-flight
calibration
\citep{moscal} did not lead to good fit results for our EQ Peg data. It turned out
that especially for the value of the core radius ($r_{c}$) binning and pile-up
effects have to be considered, while the variation of the slope $\alpha$
is only moderate. In order to find a better representation for the shape of the
PSF we carried out a parameter study for our data from EQ Peg and for
XMM-Newton observations of single point sources, e.g., $\epsilon$~Eri with a
comparable detector configuration and pile-up level. From this study we
found a slope of $\alpha=1.45$, which agrees with the calibration value,
and a value for the core radius of $r_{c}=4.45$, i.e., a reduction of
$\sim 20 \%$, to be better suited to model our data.

\begin{figure}[ht]
 \resizebox{\hsize}{!}{\includegraphics{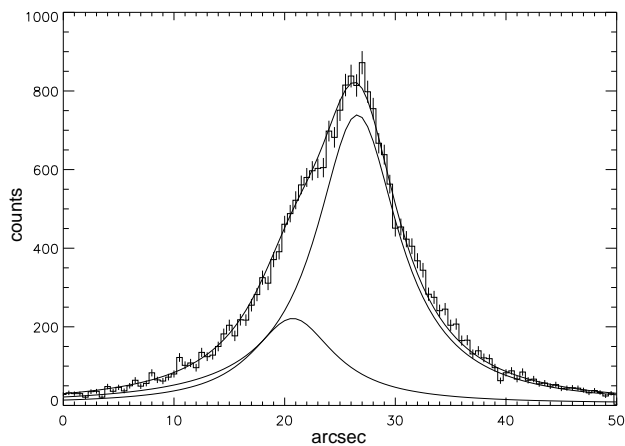}}
\caption{\label{posfit}The event distribution of EQ Peg, 
%which is a trace through the center of the emission, 
fitted with the PSF
model. Shown are the individual components as well as the sum of both
components compared to the data (histogram).}
\end{figure}

With the redetermined PSF parameters we modelled the event distribution.
In Fig. \ref{posfit} we show a
one-dimensional representation of our fit results for the EQ Peg
observation.
For this purpose we integrated along the declination axis,
which almost matches the main axis of the EQ Peg system and binned the data into a histogram.
As can be seen in Fig. \ref{posfit}, the model describes the data very well.

\section{Results}
\label{results}

\subsection{Determination of exact source positions}
\label{total}

A first inspection of Fig.~\ref{image} suggests the presence of two sources.
Since the angular resolution of the MOS1 detector is of the same order as
the separation of the two components of EQ Peg ($\sim 5\arcsec$), it is clear
that we are operating at the limit of the spatial resolving power of the MOS1
detector.
We applied our PSF algorithm to the EQ Peg dataset to
determine precise source positions. We use the MOS1 data from a
50x50\,$\arcsec$ field centered on the position of the EQ Peg system.
The calculated source positions as listed in
Table~\ref{pos} agree well within the errors
with the optical positions taken from literature \citep{perry97}. Here proper motion
corrections were not applied since they are small because the observation took place in July 2000.
We therefore conclude that we indeed identified the two X-ray sources with the
optical counterparts.

The algorithm was also used to determine the absolute
number of counts per source and the count ratio of the sources.
EQ Peg~A was on average a factor of $\sim 3.5$ brighter than EQ Peg~B during
the total observation.

\begin{table}[!ht]
\caption{\label{pos}Position fit results (1 $\sigma$ errors), ref. data from Simbad
(FK5/2000)}
{\scriptsize
\begin{tabular}{lcc|cc}\hline
& \multicolumn{2}{c}{EQ Peg A} & \multicolumn{2}{c}{EQ Peg B} \\\hline
   & Fit & Ref. & Fit & Ref. \\
   RA (23:31:)&  52.17\,$\pm$\,0.01&  52.18 & 52.57\,$\pm$\,0.02& 52.53 \\
    & (352.9674$^\circ$) & & (352.9690$^\circ$) & \\
   DEC (+19:56:)& 14.10\,$\pm$\,0.05&  14.10 & 14.05\,$\pm$\,0.10& 13.90 \\
   & (19.9373$^\circ$) & & (19.9372$^\circ$) & \\\hline
 EQ Peg System &  Fit & \multicolumn{2}{c}{Ref.}& \\
 Separation ($\arcsec$)& 6.0\,$\pm$\,0.3  & \multicolumn{2}{c}{5.2}&  \\\hline
 Count ratio A/B &\multicolumn{4}{c}{ 3.4\,$\pm$\,0.2 \ \ \ 4.2\,$\pm$\,0.3 (quies.) \ \ \  2.9\,$\pm$\,0.2 (flare)} \\
\end{tabular}
}
\end{table}

\subsection{Identification of flare activity}
\label{flare}

\begin{figure}[!ht]
\resizebox{\hsize}{!}{\rotatebox{270}{\includegraphics{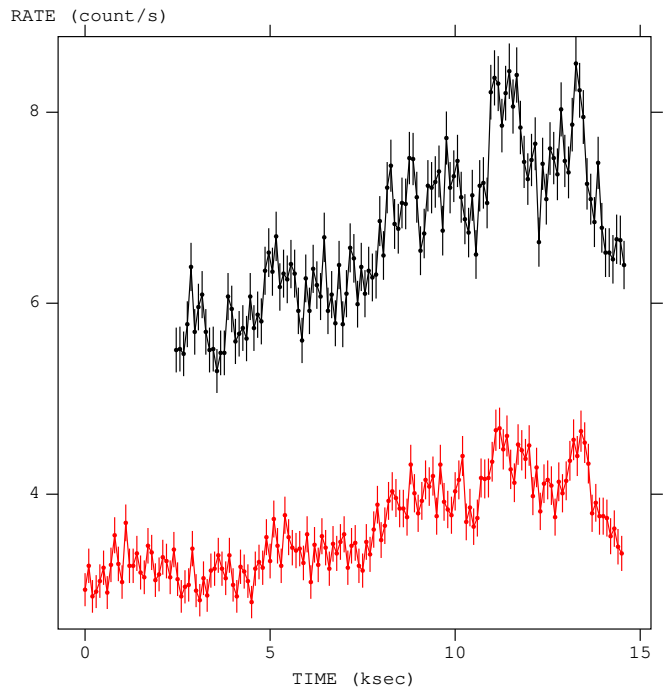}}}
\caption{\label{lc}Light curve of the EQ Peg observation as measured by the PN (black)
and MOS1+2 (grey) detectors with 100\,s binning.}
\end{figure}

From the image obtained with MOS1 the two components of EQ Peg can be
separated. 
%However, reliable light curves and spectra can only be extracted
%for the sum of the two components. 
In Fig.~\ref{lc} we plot the light curves
for the EQ Peg system as observed with the different EPIC detectors.
Inspection of the total light curves in Fig.~\ref{lc} shows that the EQ Peg system
stayed more or less quiet during the first 8\,ksec, afterwards
a rise in count rate is detected in all three detectors.
We therefore divided the data set into
two parts separated at t=7.86\,ksec where we consider the first part the
quiescent phase and the second part the flaring phase.

\begin{figure}[ht]
\hspace{-5mm}\includegraphics[width=95mm]{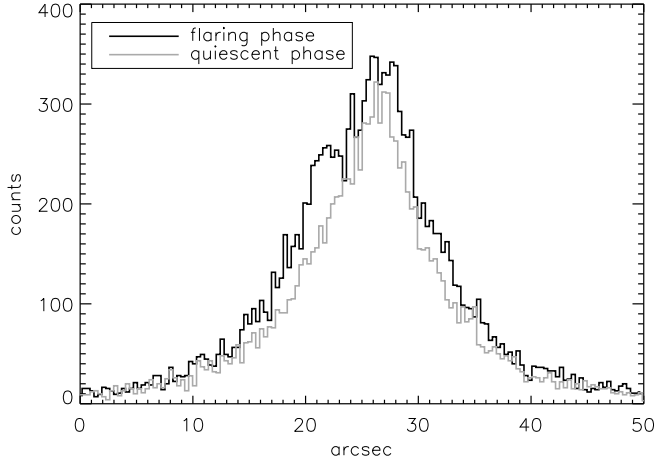}
\vskip-5.0mm
\caption{\label{evdis}Comparison of the event distribution during the quiescent and flaring phase.}
\end{figure}

In Fig.~\ref{evdis} we display the event distributions of these two subsets;
the histograms were created in the same way as in
Fig.~\ref{posfit} and are corrected for the different integration times. Here EQ
Peg~A is the X-ray brighter component on the right, EQ Peg~B corresponds to
the weaker component on the left edge of the event-distribution. In the
quiescent phase EQ Peg A dominates the emission and the second component is
only marginally visible, while during the flaring phase EQ Peg~B brightens up and
becomes more clearly visible.

The analysis of the X-ray images as carried out for the total
observation (Sect.~\ref{total}) can be repeated for the different phases of
activity.
Application of our PSF algorithm to the two subsets with variation of only
the amplitude parameters returned the count rate of EQ Peg~A to be
a factor $\gtrsim 4$ higher than for EQ Peg~B in the quiescent phase, while in
the flaring phase the ratio was $\lesssim 3$.
The results of the fitting procedure for the EQ Peg observation are summarized in Table~\ref{pos}.

\subsection{Reconstruction of individual light curves}

Having found that the major flaring activity is due to EQ Peg~B, we decided
to carry out a systematic light curve analysis of both components of the EQ Peg system.
As a first approach we extracted
two different light curves from the MOS1 data
by placing a circular region with 2.5$\arcsec$ radius around each
component, concentrating on the core of the PSF.

\begin{figure}[!ht]
 \resizebox{\hsize}{!}{\rotatebox{270}{\includegraphics{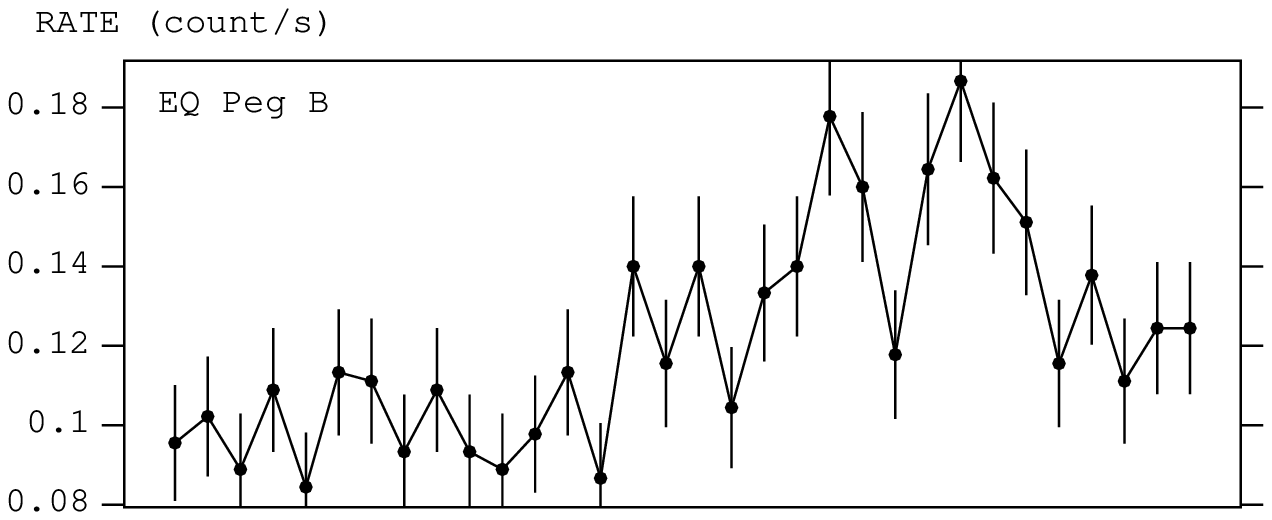}}}
\vskip-4.6mm
 \resizebox{\hsize}{!}{\rotatebox{270}{\includegraphics{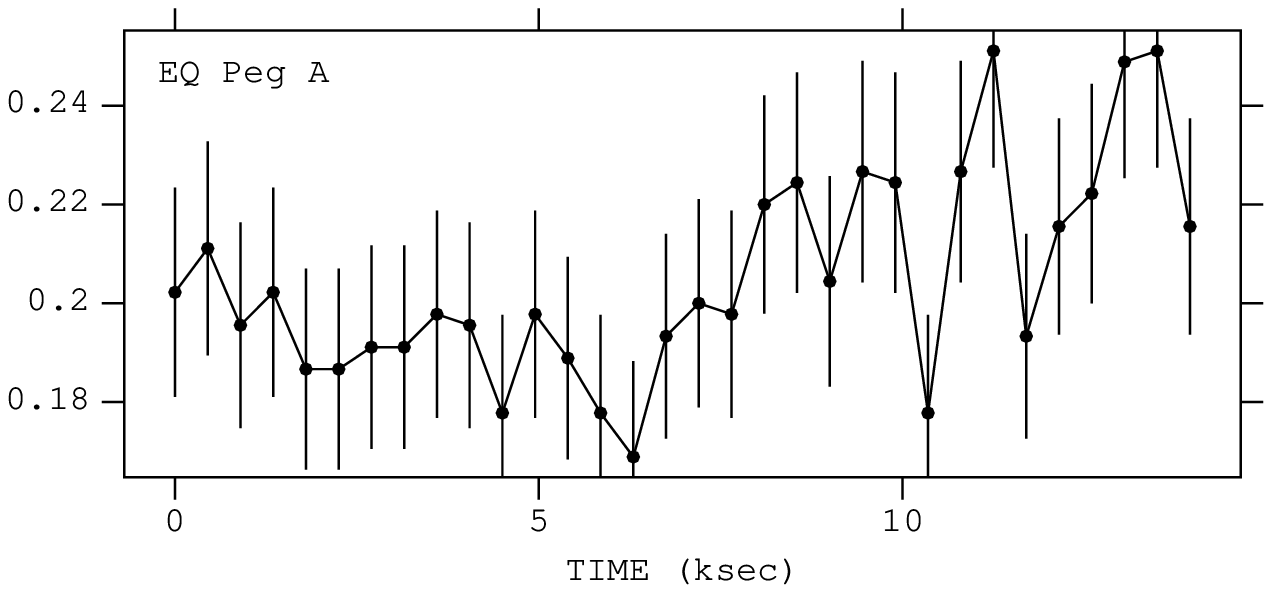}}}
\caption{\label{lcc}Light curves of the two components extracted from a circular region around each component.}
\end{figure}

In Fig. \ref{lcc} we show these light curves binned into 7.5\,min bins.
Again, EQ Peg~B is more variable and its count rate rises by a factor of
$\sim 2$ at the peak of the
flare while EQ Peg~A shows only marginal brightening compared to the quiescent
emission level. Clearly, the extraction regions used do contain photon contamination from
the respective other component due to the wings of the PSF. Nevertheless, the
individual light curves also suggest uncorrelated variability between the A and B
components. In particular, the flaring at the end of the observations
(e.g, t$\,\gtrsim 11$\,ksec) seems to originate from
the A component.

\begin{figure}[!ht]
 \resizebox{\hsize}{!}{\includegraphics{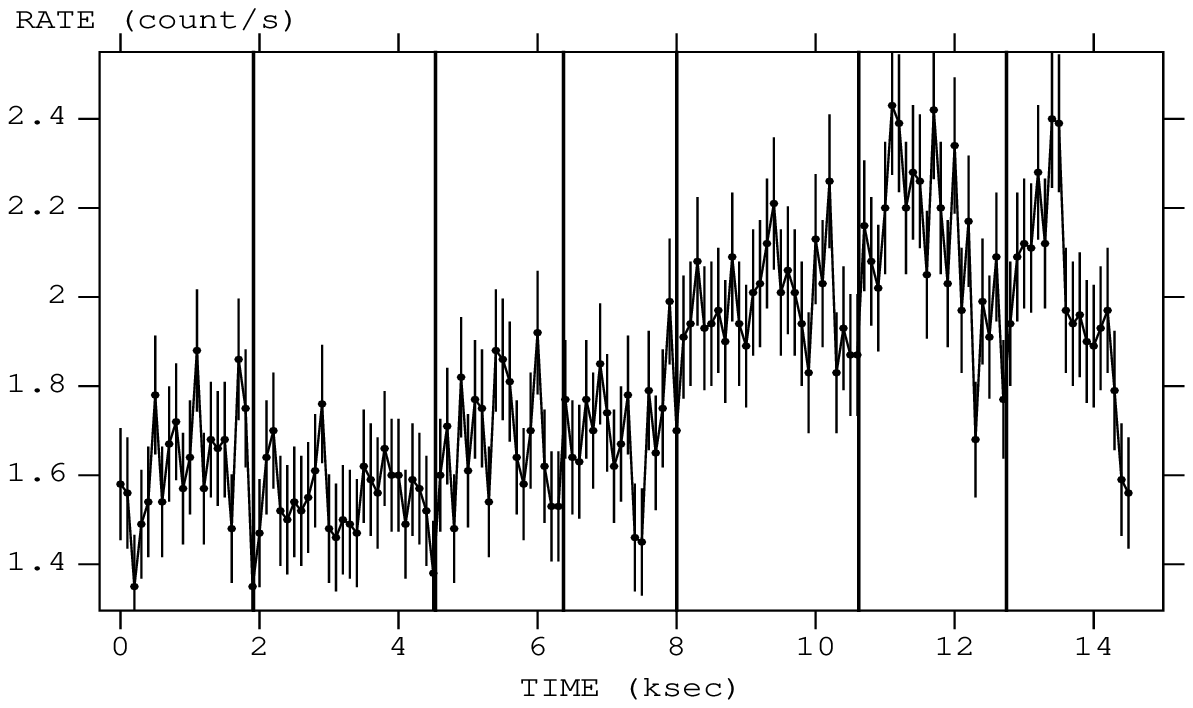}}
\vskip-35.0mm
\includegraphics[width=95mm]{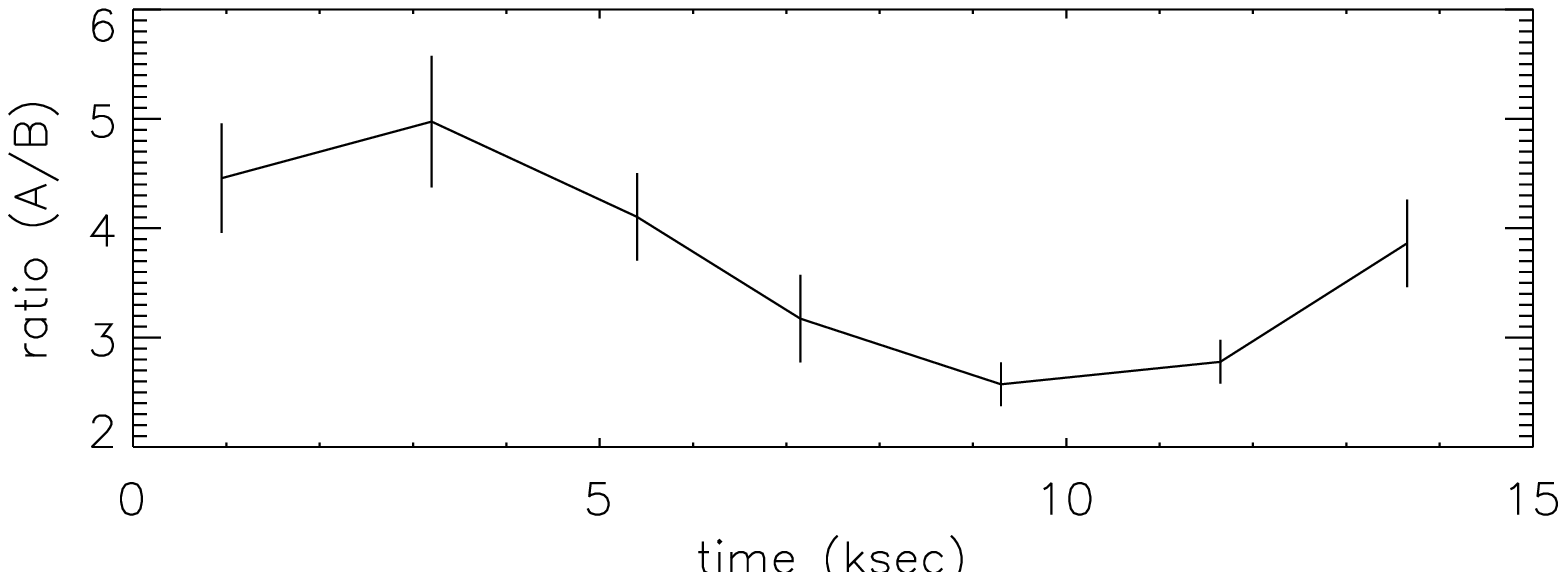}
\caption{\label{ratio}Division of the MOS1 light curve into seven time intervals (top) 
and derived flux ratios (bottom).}
\end{figure}

\begin{figure}[!ht]
\vskip-45.0mm
\includegraphics[width=95mm]{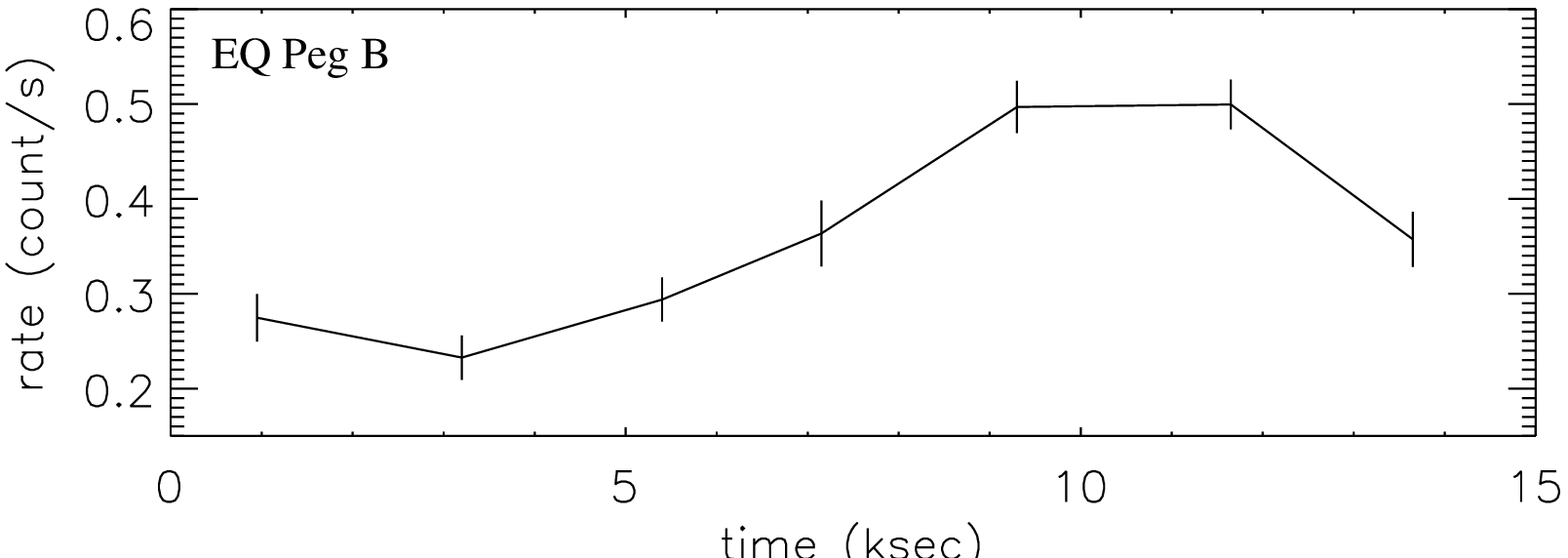}
\vskip-35.0mm
\includegraphics[width=95mm]{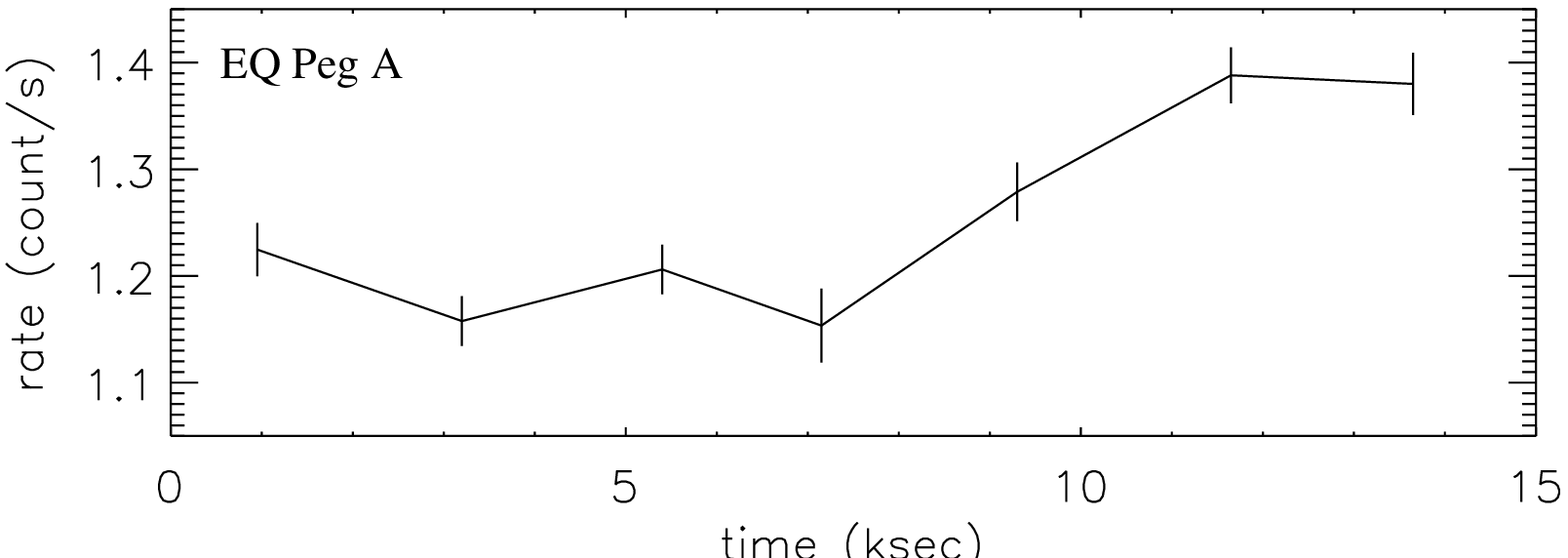}
\caption{\label{rates}Light curves of EQ Peg B (top) and A (bottom) calculated with the PSF fitting algorithm.}
\end{figure}

For a more detailed quantitative treatment we utilize our PSF algorithm in order
to reconstruct the individual light curves.
We divide the dataset into seven time intervals covering the various phases of
activity as shown in the upper panel of Fig.~\ref{ratio}. From our PSF algorithm
a count ratio for each time interval can be determined and in the bottom
panel of Fig.~\ref{ratio} we show the development of this count ratio.
While the main flaring activity is located on EQ Peg~B (indicated by the
decrease of the A/B count ratio after $\sim 5$\,ksec) there is also some
activity on EQ Peg~A especially during the later phase of the observation.

From these ratios we calculated light curves for each individual component,
which are shown in Fig. \ref{rates}. Although the light curves consist of rather
large time bins the main features visible in Fig.~\ref{lcc} are also
present, i.e., a flare on EQ Peg~B around 10\,ksec and the flaring
activity on EQ Peg~A towards the end of the observation. The rise in count rate
associated with the flare activity on both components is nearly equally strong,
i.e., $\sim 0.2$\,counts/s, however, the relative change in count rate is much
higher on EQ Peg~B.

\subsection{Development of the spectral hardness}

In order to quantify eventual changes in the physical conditions accompanied by
rising count rates we calculate a spectral hardness ratio for the sum of both components in
two energy bands, resp. $0.2-0.5$\,keV (soft) and $1.0-10.0$\,keV (hard). The
hardness ratio was calculated from PN data which were cleared for pile-up effects.

\begin{figure}[!ht]
 \resizebox{\hsize}{!}{\rotatebox{270}{\includegraphics{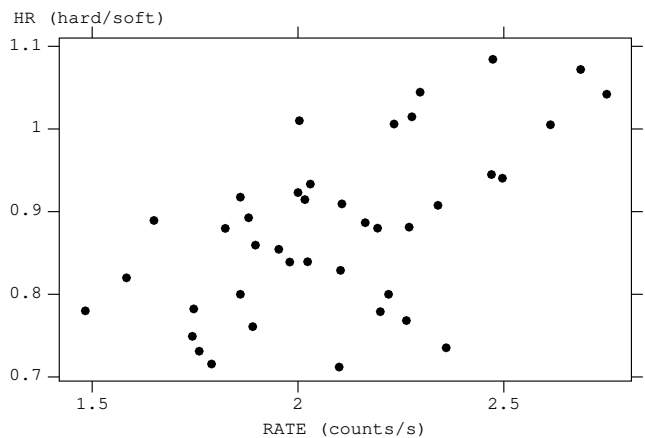}}}
\caption{\label{hr}Hardness ratio (1-10\,keV/0.2-0.5\,keV) derived from PN data vs. count rate.}
\end{figure}

In Fig.~\ref{hr} we show the spectral hardness ratio (hard/soft) 
for this observation binned every five minutes vs. the measured count rate. 
The hardness ratio increases during the times where we also 
observe increases in count rate. We calculated a linear correlation coefficient and found a correlation probability of 99.9\,\%.
We therefore interpret the simultaneous increase in count rate and hardness of the emission
as flare heating of the coronal plasma.
When comparing the calculated average hardness ratios during the time intervals defined in
Sect.~\ref{flare} as the quiescent phase and the flaring phase 
we find an increase in hardness by 12$\pm$2\,\% from the quiescent phase into the flaring 
phase, whereas the increase during individual flares is substantially stronger.

The average X-ray luminosity between 0.2\,keV and 10.0\,keV was calculated
to be $4.2\times10^{28}$\,ergs/s from application of spectral models.
We compared the X-ray luminosities obtained from a number of different reasonable 
models and find consistent results.
In comparison with previous measurements the energy flux during this
observation was $\sim 20\%$ above the flux obtained with ROSAT
\citep{huensch99} and $\sim 50\%$ below the values obtained with
{\it Einstein} \citep{vai81} in the respective energy bands.

\section{Summary and Discussion}
\label{summ}
Our analysis of EQ Peg is another example of how
the high-resolution X-ray telescopes {\it Chandra} and XMM-Newton 
allow to resolve sources down to a unprecedented spatial resolution
(for other examples see \cite{stelzer03} and \cite{au03}).

With the XMM-Newton observation of the EQ Peg system we were able to separate
the two components for the first time in X-rays. Using a PSF model fit procedure
we can reconstruct the source positions and show that both
components are flaring X-ray emitters. On average, we found the
A component brighter by a factor $\sim 3.5$ for the total observation.

During this observation a series of medium flares was detected. We were able
to determine count ratios for EQ Peg~A/B for the different phases of activity.
During the early (quiescent) phase of the observation the emission is strongly
dominated by EQ Peg~A, which is a factor of $\sim 4-5$ brighter than EQ Peg~B.
Comparison of the quiescent and active phases made it possible to associate most of the
flaring with EQ Peg~B, which nearly doubled it's X-ray brightness
during the peak of the flare. The count ratio during the peak of the flare on
EQ Peg~B dropped to $\sim 2.5-3$. We also found evidence for flaring activity
on EQ Peg~A towards the end of the observation, consistent with
previous findings that both stars exhibit flaring behavior
\citep[e.g.,][]{rod78}. In fact, the relative brightening during the flares is
much stronger for EQ Peg~B, but the absolute increase in flux is comparable for
both stars. The energy released by these flares is obviously very similar,
although the quiescent emission level is quite different.
The flaring X-ray emission of the EQ Peg system shows the typical hardening
in the spectral energy distribution as expected for stellar flares.

\begin{acknowledgements}
This work is based on observations obtained with XMM-Newton, an ESA science
mission with instruments and contributions directly funded by ESA Member
States and the USA (NASA).\\
J.R. and J.-U.N. acknowledge support from DLR under 50OR0105.

\end{acknowledgements}

\end{document}